# Pinpointing the desert of the Ruoqiang County, Western China

Amelia Carolina Sparavigna
Department of Applied Science and Technology, Politecnico di Torino, Italy

*A man-made texture on the desert soil of a county of the Western China is visible in satellite images, more than 8 kilometers long and 50 meters wide. This seems to be the result of a detailed geophysical survey of the region that led to the discovery of a large nickel ore. Therefore, the analysis of the satellite imagery, performed to find such textures created by the sampling of soils, can help anticipating information on the economical potentialities of a site.*

In [1], I proposed a discussion on the Loulan Kingdom, an ancient kingdom on the Silk Road, using the satellite maps to illustrate its location at the East end of the Taklamakan desert. The region was recently investigated by some archaeological expeditions [2], aimed to determine the precise location of towns and related communities of the Luolan and Shanshan states [3-5] in the Lop Nor region. The archaeological researches [2] found a new site and investigated the radiocarbon ages of Loulan, LE, Qieerqiduke and other sites.

It is interesting to remark that in several satellite maps it is possible to find some remains of the walls of these towns. One in particular, that of LE, is quite visible in the maps [1]. However, one of the sites discussed in [2] cannot be seen in the freely available satellite maps; it is that of the Qieerqiduke City (38°57'10.73"N, 88°10'4.61"E), about 6 km southeast from Ruoqiang County [6,7], in Xinjiang, Western China. Archaeologists report that some residual earthworks are visible. The Ruoqiang county, also known as Charkliq, was a kingdom that, from the 1st century BCE became part of the Kingdom of Loulan [6,7]. An explorer and archaeologist, Aurel Stein, visited the oasis of Charkliq in 1906 [8], founding a little village ruling a very large desert district, including the salt lake of Lop Nor.

Near the archaeological site of Qieerqiduke, in the satellite maps, we can see a man-made texture on the soil, a huge band which seems created by relatively small holes or mounds (Fig.1). It is partially a band and then becomes composed of squares as in chessboard rows. The overall band is more than 8 kilometers long and more than 50 meters large. This curious texture on the desert soil was probably produced by the pinpointing of geophysical researches. An alternative interpretation could be that of a structure for stabilizing the soil, but there is no evidence supporting it. Evidence for a survey origin exists: the China Daily has recently announced that some geological workers have discovered a large nickel ore in Ruoqiang County. According to [9], this ore will help the request of industrial nickel resources. The geological departments and mining companies carried out in all the Ruoqiang County a four-year exploration starting from 2009. It seems that the surveyed zone was quite large.

The Xinjiang nickel is located in one of the magma belts of the Talimu basin igneous rocks [10], where there are many mafic-ultramafic intrusions associated with magmatic Cu-Ni deposits. Besides nickel and copper deposits, the Lop Nor is rich of potassium salts that China will move on a new railway, the construction of which ended in September 2012, to make fertilizers [11]. The Lop Nor was known as the "sea of death" for its high salt content. This lake, fundamental for the ancient kingdoms on the Silk Road, was the largest lake in northwest China before it dried up in 1972.

As explained in [12], different techniques are used in searching for different minerals, but the principles are similar. There is a first survey across a large area. In the case that an anomaly is found, a detailed survey is carried out to find the origin of this anomaly. A series of trenches or boreholes is used to find the source of the anomaly. Usually, the prospecting geologist works with a mining geologist, in excavating pits for evaluating the ore. The survey can be made on "green field" regions, where no ore deposits have previously been found, or on "brown field" areas, near known deposits. In the case of a prospection of a green field area, remote sensing data for the region are examined to find unusual features. The remote survey includes satellite data using visible light or other parts of the spectrum, and aerial photographs. The next stage might be a mapping of gravity and geomagnetism. Geomagnetic ground survey provides details on magnetic anomalies; an electrical surveying might find conducting material deposits. The sampling is the last stage of the survey, normally on a grid pattern, to pinpoint the source of the anomaly [12].

Let us choose the coordinates 38.960869,88.159586 of the Ruoqiang County. In the Google Maps we see what could be the beginning of a mining activity. In the Bing or Nokia Maps, which are older, this structure does not exist; however, we see the pinpointing of the area. As a matter of fact, the last part of the survey, the sampling, can be visualized in the satellite maps, revealing that the site resulted with some potentialities after the previous analyses. We have therefore a feedback on the same satellite images, used for preliminary investigations, of further localized activities.

As told in [1], the satellites can be used to find some textures, different from the background, which can reveal some archaeological remains of the past. Of course, the same analysis can provide information on the future of a region. If we see during the analysis of satellite imagery some textures created by the sampling of the soil, we can argue that the site has economical potentialities. Then the image analysis can help anticipating information on the economical future of the local area and its connections on a larger scale.

**References**
[1] A.C. Sparavigna, The road to the Loulan Kingdom, arXiv, 2012, http://arxiv.org/abs/1210.5702
[2] Lü HouYuan, Xia XunCheng, Liu JiaQi, Qin XiaoGuang, Wang FuBao, Yidilisi Abuduresule, Zhou LiPing, Mu GuiJin, Jiao YingXin, Li JingZhi, A preliminary study of chronology for a newly-discovered ancient city and five archaeological sites in Lop Nor, China, Chinese Sci Bull, 2010, Vol.55, pag.63–71.
[3] Wikipedia, Loulan Kingdom, http://en.wikipedia.org/wiki/Loulan_Kingdom
[4] Yuan Guoying, Zhao Ziyun, Relationship between the rise and decline of ancient Loulan town and environmental changes, Chinese Geographical Science, 1999, Vol.9, Number 1, pag.78-82.
[5] Wikipedia, Shanshan, http://en.wikipedia.org/wiki/Shanshan
[6] Wikipedia, http://en.wikipedia.org/wiki/Kingdom_of_Charklik
[7] Wikipedia, http://en.wikipedia.org/wiki/Ruoqiang_County
[8] 'Serindia', Sir Aurel Stein, 1921, http://www.centralasiatraveler.com/cn/xj/cq/ch-viii-charchan_serindia_stein.html
[9] Large nickel ore discovered in China's Xinjiang, China Daily, 6 October 2012, http://www.chinadaily.com.cn/business/2012-10/06/content_15798222.htm
[10] JIA Hong-xu, LAI Tao,WANG Hen, TAN Ke-Bin, LU Hong-Fei, The geological characterization of Hongshishan Nickel deposit in Ruoqiang, Xinjaing and its prospecting indications, Xinjiang Geology, 2011, No.6 Geological party of Xinjiang Bureau of Geology and Mineral Resources, Hami, Xinjiang, China, http://en.cnki.com.cn/Article_en/CJFDTOTAL-XJDI201101015.htm
[11] Railway traverses 'sea of death', China Daily, 22 July 2012, http://www.chinadaily.com.cn/china/2012-07/22/content_15606622.htm
[12] Chris King, The planet we live on - the beginnings of the Earth Sciences, 2010,

http://www.learndev.org/dl/Science/EarthScience/ThePlanetWeLiveOn-FrontMatter.pdf

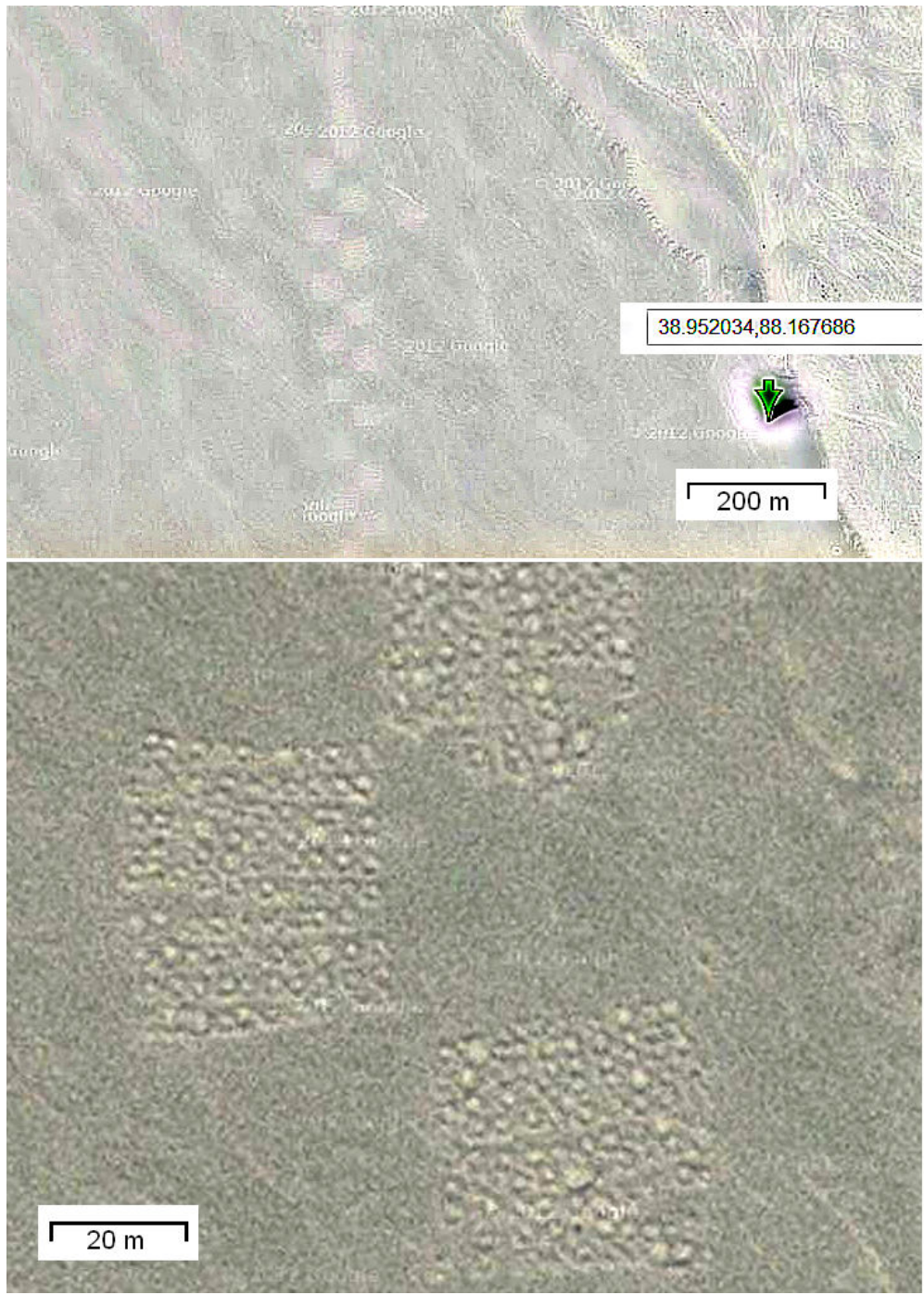

Fig.1 In these Google Maps, we see a faint structure which looks like a band of dots, made by the pinpointing of geological surveying in the Ruoqiang County. Enhancing the images, we find a band of dots which is more than 8 kilometers long and composed of 40 meters wide grids.